\documentstyle[12pt]{article}
\input{psfig}
\begin{document}
\begin{titlepage}
\begin{center}
\hfill hep-th/0007110\\
\hfill IP/BBSR/2000-27\\
\vskip .2in

{\Large \bf Triality of Four Dimensional Strings and Networks}
\vskip 0.5in
{\bf Alok Kumar\footnote{e-mail: kumar@iopb.res.in},
Aalok Misra\footnote{e-mail: aalok@iopb.res.in},\\
Institute of physics,\\
Bhubaneswar 751 005, India}
\vskip 0.5 true in
\end{center}

\begin{abstract}
We apply the string-triality\cite{duffetal} 
to argue the existence of the string-networks of solitonic 
$T$ and $U$-strings in the heterotic theory on $T^6$, 
$S$ and $T$-strings in IIB on $K3\times T^2$ and, $S$ and
$U$-strings in IIA on $K3\times T^2$.
We then show the existence of the above 
heterotic string networks by analyzing 
the supersymmetry property of the supergravity solutions. 
The consistency of these networks with 
supersymmetry in the case of IIA and IIB theories is also 
argued. Our results therefore give further evidence in favor
of the string-triality in  four dimensions. 
\end{abstract}
\end{titlepage}
\newpage

Heterotic string theory on $T^4$ is related by string/string
duality to type IIA theory on $K3$ \cite{Luloop}-\cite{Harvey}.
 This also leads to 
a duality among these theories, upon their compactifications
to four dimensions \cite{duffetal}.
 This duality interchanges the axion-dilaton 
modulus of one theory with the K\"{a}hler modulus of 
the dual theory. 
Combining it with mirror symmetry (that interchanges the 
K\"{a}hler modulus with the complex structure modulus)\cite{Dine}
leads to string/string/string triality \cite{duffetal}
relating types IIA, IIB
theories on $K3\times T^2$ and heterotic string theory on $T^4\times
T^2$. Let ${\cal T}_{XYZ},\ {\cal T}\equiv H,A,B$ 
denote any of the three theories with
$S\equiv$ the axion-dilaton modulus, $T\equiv$ K\"{a}hler
modulus and $U\equiv$ complex structure modulus. Then
mirror symmetry takes $H_{STU}$ to $H_{SUT}$. 

It is known that the equations of motion of the heterotic strings
admits a solution called an
$S$-string\cite{duffetal,Duffstrong,kehagias}. 
Further, string-string duality takes $H_{STU}$ to $A_{TSU}$, which under
mirror symmetry goes to $B_{TUS}$. Then $A_{TSU}$ and
$B_{TUS}$ admit a solution called a $T$-string
\cite{duffetal,Duffstrong,kehagias} 
Finally, under string-string duality, $B_{TUS}$ goes 
to $B_{UTS}$, which under
mirror symmetry goes to $A_{UST}$. Accordingly, 
equations of motion of $B_{UTS}$
and $A_{UST}$ admit a solution called a $U$-string. 
More precisely, string-triality requires the following 
relations among various string solutions: $S$/$T$-strings
of the heterotic theory are mapped to $T$/$S$-strings in IIA 
(on $K3\times T^2$), with $U$-string mapped to itself. Similarly, 
$S$/$U$-strings of the heterotic theory are mapped to 
$U$/$S$-strings in IIB on $K3\times T^2$ and $T$-string 
maps to itself. Finally, $T$/$U$-strings of IIA (on 
$K3\times T^2$) are mapped to $U$/$T$-strings in IIB, 
with $S$-string going to an $S$-string of IIB.
The string tensions of ($S$, $T$
$U$)-strings in the heterotic, IIA and IIB theories are expected to
map into each other under the above string-dualities.

These statements carry over to the string-networks
\cite{sen97} - \cite{zwieb} as well. To find
out the kind of string-networks that exist in each of these theories, 
we start with the IIB case. In this case, we argue that 
string-network of `$(p, q)$ $S$-strings' exists in IIB on 
$K3\times T^2$. To construct the $(p, q)$-multiplet of such
strings, we notice that 
the $S$-string in IIB is the dimensionally reduced ten-dimensional 
F-string\footnote{Strictly speaking, this is a `smeared' or 
delocalized string solution, independent of the coordinates 
along $K3\times T^2$.}. 
This string couples, in four dimensions, to the
dilaton and axion fields of the compactified IIB theory. 
These fields, in turn, are 
the dimensional reduced version of the dilaton and NS-NS sector
2-form fields in ten dimensions. Now, since the $SL(2, Z)$
S-duality of type IIB is preserved under geometric 
compactifications, such as $K3\times T^2$, one can form 
an $SL(2, Z)$ multiplet of such S-strings using their 
connection with the F-string of IIB. 
The argument supporting the
existence of such a string-network in the type IIB side then 
predicts the existence of the network of `$U$-strings' in 
the heterotic theory. These strings couple to scalars, 
identified as the $U$-moduli on $T^2$ formed out of
metric components on $T^2$ after removing a determinant factor.
More precisely, the metric on $T^2$ is written as:
\begin{eqnarray}
\label{eq:T2metric}
G \equiv e^{\rho - \sigma}\pmatrix{e^{-2\rho} + c^2 & -c \cr
 -c & 1},
\end{eqnarray}
then $U\equiv c + i e^{-\rho}$ \cite{duffetal}. 
Then using IIA/Heterotic duality, where $U$-moduli of one of these theories
are mapped to that of the other one,
we observe that `$U$-string' networks 
should exist in type IIA theories as well. Furthermore, using
the mapping between IIA and IIB, where $T$ and $U$ are interchanged,
with $T \equiv B_{45} + i \sqrt{det (G_{ab})}$, 
one finds that a network of $T$-strings should exist in type IIB 
on $K3\times T^2$. This using IIB/heterotic duality
then implies a network of $T$-strings 
in the heterotic case. Finally, using IIA/heterotic duality one
has a network of $S$-strings in type IIA theories. 
To summarize, the string/string/string-triality in four dimensions
implies string-networks of the $(p, q)$-multiplets of
$S$ and $T$-strings in type IIB theories, $T$ and $U$-strings
in the heterotic theories and $S$ and $U$-strings in type IIA 
theories. In this paper, we will verify that such string-networks
indeed exist in these theories. We show this first for the hetetotic 
theory explicitly. 
The existence of string networks in other cases
is argued on the basis of their constructions in higher 
dimensions.

We now start with a review of the string solutions of the 
heterotic theory. $S$-string solution is given by
\footnote{Modification of $S, T, U$-string solutions to obtain finite
energy solutions, by using modular invariant functions, is given in 
\cite{kehagias}. However, since this solution reduces to the ones
given below in appropriate limit, the supersymmetry condition from these 
solutions \cite{kehagias} is expected to be consistent with
the ones written in this paper.}\cite{duffetal}
\begin{eqnarray}
\label{eq:Sdef}
ds^2 & =e^{\eta}(-(dt)^2+(dx^1)^2) + (dr^2+r^2d\theta^2),\cr
& \cr
& S = S_1 + i S_2  = a + i e^{-\eta} = {1\over 2\pi i} ln{z\over r_0}, 
\end{eqnarray}
where $z=x_2+ix_3$ ($x_{2,3}$ being the coordinates transverse to the
string world-sheet coordinates $x_{0,1}$), 
`$a$' is the axion field in four dimensions defined in terms
of the $2$-form field $B_{\mu \nu}$ as: 
$\epsilon^{\mu\nu\rho\lambda}\partial_\lambda a
=\sqrt{-g}e^{-\eta}g^{\mu\nu_1}g^{\nu\nu_1}
g^{\rho\rho_1}H_{\mu_1\nu_1\rho_1}$. $\eta$ is the 
$D=4$ dilaton. $T$-string solution is given by:
\begin{eqnarray}
\label{eq:Tdef}
ds^2& =-(dt)^2 + (dx^1)^2 + e^{-\sigma(r)}(dr^2+r^2d\theta^2)\cr
& \cr
& T  = b + i e^{-\sigma} = {1\over 2\pi i} ln{z\over r_0},
\end{eqnarray}
where $b$ and $\sigma$ are defined as:
$T = B_{4 5} + i \sqrt{det(G_{ab})} \equiv b + i e^{-\sigma}$.
The $U$-string solution is given by:
\begin{eqnarray}
\label{eq:Udef}
ds^2 & =-(dt)^2+(dx^1)^2+e^{-\rho(r)}(dr^2+r^2d\theta^2),\cr
& \cr
& U = c + i e^{-\rho} = {1\over 2\pi i} ln{z\over r_0}.
\end{eqnarray}
Solutions in II A/B theories can be obtained from these by string-triality.
We refer to the soultions in eqns.(\ref{eq:Sdef}), 
(\ref{eq:Tdef}), (\ref{eq:Udef}) as 
$(1, 0)$ $S$-string, $(1, 0)$ $T$-string and 
$(1, 0)$ $U$-string respectively. More general solutions,
referred as $(p, q)$ $S$, $T$, $U$-strings can be generated from 
these by applying appropriate $SL(2, Z)$ tranformations. 
We refer to them as $SL(2, Z)_H$. 
These $SL(2, Z)$  transformations 
in turn are the mappings of the $SL(2, Z)_B$ $S$-duality 
transformations of the ten-dimensional type IIB theories. 
More precisely, the $SL(2, Z)$ of IIB in ten-dimensions
can be identified as a group of constant coordinate tranformation
involving any two of the six compact coordinates in the heterotic
theory. All these $SL(2, Z)$'s are in fact part of the 
$O(6, 22, Z)$ $T$-duality symmetry of the heterotic string,
belonging to its diagonal subgroup $SL(6, Z)$. 
The mapping from 
$SL(2, Z)_B$ to $SL(2, Z)_H$ is obtained by 
identifying their actions on 
various fields in IIB and heterotic theories. 
in four dimensions. Field content of the heterotic theory 
has a metric ($g_{\mu \nu}$), scalars:  $S$-modulus defined earlier,
28 vector fields transforming in a vector representation of
$O(6, 22)$ and 132 additional scalars forming a coset: 
$O(6, 22)/O(6)\times O(22)$. To identify the appropriate
$SL(2,Z)$, we denote indices $(0,1,2,3)$ as $(\mu, \nu)$. 
Internal indices are denoted as: 
$a\equiv (4, 5)$,  $i\equiv (6,..9)$, $ m=(7, 8, 9)$.
In the IIB case below, we will denote the internal 
$K3$ by indices $(\bar{I}, \bar{J})$ etc..

In the IIB theory, the fields corresponding to the heterotic theory
mentioned in the last paragraph are obtained from the $K3\times T^2$
compactification of the ten dimensional fields. 
The metric ($g_{\mu \nu}$) comes from the 
components of the ten-dimensional 
metric. 2 gauge fields ($A_{\mu}$'s) are the Kaluza-Klein modes on $T^2$:
$g_{\mu a}, B^1_{\mu a}, B^2_{\mu a}$. Remaining 22 gauge fields
($A_{\mu}$'s) arise from the self-dual 4-form in ten dimensions:
3 come from the self-dual and 19 from the anti-self-dual
2-forms on $K3$. Among scalars, 2 are the 
ten-dimensional axion-dilaton fields, 3 come from the 
components of metric along $T^2$, and 58 scalars are 
the geometric moduli of $K3$. The 2-form fields give scalars
from the components along $T^2$: $(B^1_{a b}, B^2_{a b})$ as well
as from the space-tiem components $(B^1_{\mu\nu}, B^2_{\mu\nu})$. 
They also give rise to 22 scalars each from $B^1_{\bar{I} \bar{J}}$, 
$B^2_{\bar{I}\bar{J}}$. The 4-form field gives another 23 scalars, out of 
which 22 come from the 2-forms on $K3$ and one from 
0-form. Combining these one gets an identical 
field content as in the heterotic theory.

To obtain a precise connection between $SL(2, Z)_B$ and 
$SL(2, Z)_H$,
we now discuss the tranformation properties of above fields,
both in IIB and heterotic theories.  In the IIB case,
4 gauge fields arising as $\pmatrix{B^1_{\mu a}$, \\$B^2_{\mu a}}$
tranform as a vector under $SL(2, Z)_B$. The remaining ones,
coming from the ten-dimensional metric: 
($g_{\mu a}$) and the self-dual $4$-form field, are neutral under this 
$SL(2, Z)$. In the heterotic theory, since the $SL(2, Z)$
is realized as a coordinate transformation on any
two of the compact coordinates\footnote{leaving the combination
$(x^4, x^5)$, which is already used in defining the $T^2$ mentioned
above.}
mentioned above (for definiteness we choose them as $x^5$ and 
$x^6$), four gauge fields transforming as a vector
under this $SL(2, Z)$ are identified as:
\begin{eqnarray}
\label{eq:SL2vecs1}
\pmatrix{g_{\mu 5}\cr g_{\mu 6}},\>\>
\pmatrix{B_{\mu 5}\cr B_{\mu 6}}. 
\end{eqnarray}
It is evident that the remaining gauge fields:
$A^I_{\mu}, g_{\mu 4}, B_{\mu 4}, g_{\mu m}, B_{\mu m}$
are neutral under this $SL(2, Z)$ in the heterotic theory as well. 
As a result the action of $SL(2, Z)_B$ and $SL(2, Z)_H$ are
identical on vector fields in two theories.

We now compare the actions of the two SL(2, Z)'s on scalars 
in these theories. In the type IIB case, the axion-dilaton 
combination forms a coset $SL(2)/SO(2)$. In the heterotic 
theory, this role is played by the complex structure  associated
with directions $(x^5, x^6)$.  
In addition, in the IIB theory in $D=4$, 
one also has scalar fields  transforming in vector representation 
of $SL(2, Z)_B$. These scalars are:
\begin{eqnarray}
\label{eq:scalars}
\pmatrix{B^1_{a b}\cr B^2_{a b}},\>\>\> 
\pmatrix{B^1_{\bar{I}\bar{J}}\cr B^2_{\bar{I}\bar{J}}},\>\>\>
\pmatrix{B^1_{\mu\nu}\cr B^2_{\mu\nu}}.
\end{eqnarray}
The remaining scalars: $g_{a b}$, $g_{I J}$, as well as the ones
coming from the self-dual 4-form, are invariant under $SL(2, Z)$. 
In the heterotic theory, the role of the scalars transforming as
a vector (analogous to the ones in (\ref{eq:scalars}))
is played by field components:
\begin{eqnarray}
\pmatrix{g_{m 5}\cr g_{m 6}}, \>\>\>
\pmatrix{B_{m 5}\cr B_{m 6}}, \>\>\>
\pmatrix{A^I_5 \cr A^I_6},\>\>\>
\pmatrix{g_{4 5}\cr g_{4 6}}, \>\>\>
\pmatrix{B_{4 5}\cr B_{4 6}},
\end{eqnarray}
and the invariant scalars are: 
$det(G), B_{4 5}, g_{m n}, B_{m n}, g_{m 4}, B_{m 4}, A^I_m, A^I_4,
\phi, B_{\mu\nu},$ 
$g_{44}$, where $G$ is the metric in the $(x^5, x^6)$-space. 
Their numbers match exactly with the ones in IIB theory.

We have therefore obtained the mapping of the $SL(2, Z)$ 
$S$-duality transformation of IIB theory to the heterotic one. 
The $SL(2, Z)_H$ symmetry of the heterotic theory can now, in
principle, be used 
for  generating of the $(p, q)$-type solution for 
$S$, $T$ or $U$-strings starting from the $(1, 0)$ solutions 
that we wrote earlier in
eqns.(\ref{eq:Udef})-(\ref{eq:Tdef}). 
As classical solutions,
these are equivalent, related by coordinate transformations.
However, each one of them play a different role in a
network construction that we discuss below, due to 
the differences in their Killing spinor conditions.
Our aim will be to write down the conditions satisfied by the 
Killing spinors for the classical $(p, q)$-string 
backgrounds generated in this manner. 
For this we will follow the strategy in 
\cite{ortin,sen97,bhatt} to write down the supersymmetry of 
the $(1, 0)$ solution first, and then  obtain the 
supersymmetry property of the $(p, q)$-solution by 
applying $SL(2, Z)_H$ transformation. We will find some 
crucial differences among the three cases, with 
important implications for their network constructions.

We now start by
explicitly evaluating the killing spinor equations in the 
$S$, $T$, $U$-string backgrounds (\ref{eq:Sdef})-(\ref{eq:Tdef}). 
The supersymmetric transformations 
of the gravitino $\psi_\mu$, dilatino $\lambda$ and gaugino $\chi$,
(in the absence of background gauge fields)  for D=4 heterotic string theory
is given by \cite{youm,duffetal}
\begin{eqnarray}
\label{eq:susytrgen}
& & \delta\psi_\mu=\biggl[\bigtriangledown_\mu
-{1\over8}H_{\mu\nu\rho}\Gamma^{\nu\rho} +
{1\over4}Q_\mu^{\hat{m}\hat{n}}\Gamma^{\hat{m}\hat{n}}\biggr]\epsilon,
\nonumber\\
& &
\delta\lambda={1\over{4\sqrt{2}}}\biggl[\gamma^\mu{\partial_\mu(S_2 -
i \gamma^{\hat{5}}S_1)\over S_2}
\biggr]\epsilon=0,\nonumber\\
& &
\delta\chi^{\hat{m}}={1\over\sqrt{2}}
\gamma^\mu P_\mu^{\hat{m}\hat{n}}\Gamma^{\hat{n}}\epsilon,
\end{eqnarray}
where $\gamma^{\hat{5}}
\equiv{1\over4!}\epsilon_{\hat{\mu}\hat{\nu}\hat{\rho}\hat{\lambda}}
\gamma^{\hat{\mu}}\gamma^{\hat{\nu}}\gamma^{\hat{\rho}}\gamma^{\hat{\lambda}}$,
 $Q_\mu^{\hat{m}\hat{n}}\equiv (V_R L \partial_\mu
V_R^T)^{\hat{m}\hat{n}}$ 
and $P_\mu^{\hat{m}\hat{n}}\equiv (V_LL\partial_\mu V_R^T)^{\hat{m}\hat{n}}$.
$V_L$ and $V_R$ are defined using O(6,22) matrix of moduli 
$M:M=V^TV=V_R^TV_R+V_L^TV_L={1\over2}(M+L)+{1\over2}(M-L)$ where $L$
is an O(6,22)-invariant metric. Then 
\begin{equation}
\label{eq:VVLR}
V=\left(\begin{array}{c}
V_R \\
V_L
\end{array}\right)={1\over\sqrt{2}}\left(\begin{array} {cc}
E^{-1} &  E^T \\
E^{-1}& -E^T
\end{array}\right).
\end{equation}
 Hatted/unhatted Greek letters denote
four-dimensional tangent space/curved space indices, and 
Hatted/unhatted Latin letters denote six-dimensional 
tangent space/curved space indices. 
The covariant derivative in (\ref{eq:susytrgen}) is given by:
\begin{equation}
\label{eq:covdrvdef}
\bigtriangledown_\mu=\partial_\mu+{1\over4}\omega_\mu^{\hat{\mu}\hat{\nu}}
\Gamma_{\hat{\mu}\hat{\nu}},
\end{equation}
with $\omega_\mu^{\hat{\mu}\hat{\nu}}$ being the spin connection. 
We now obtain the supersymmetry conditins for `$(1, 0)$' 
$S$, $T$ and $U$-strings. 

\vskip 0.2 true in

(I) \underline{(1,0) $S$-string}
\vskip 0.2 true in

The (1.0) $S$-string solution is given in equation (\ref{eq:Sdef}). 
We now evaluate the supersymmetric variations for this background.
One can now rewrite  
\begin{equation}
\label{eq:HGammada}
-{1\over8}H_{\mu\nu\rho}\Gamma^{\nu\rho}=
-{1\over{8S_2^2\sqrt{-g}}}\gamma_{\hat{5}}^{\hat{r}\hat{\theta}}
\Gamma_{\mu\nu}\partial^\nu a,  
\end{equation}
where $\gamma_{\hat{5}}^{\hat{r}\hat{\theta}}=\gamma_{\hat{0}}\gamma_{\hat{1}}\gamma_{\hat{r}}\gamma_{\hat{\theta}}$
The spin connection $\omega_\mu^{\hat{\mu}\hat{\nu}}$ for the S-string is
given by:
\begin{equation}
\label{eq:spconSst}
\omega_\mu^{\hat{\mu}\hat{\nu}}=\left(\begin{array}{cccc} \\
0 & 0 &
{\delta_\mu^0(\partial_r\eta(r))e^{\eta(r)/2}\over 2} & 0 
\\
0 & 0 &
{\delta_\mu^1(\partial_r\eta(r))e^{\eta(r)/2}\over 2} & 0 
\\
- {\delta_\mu^0(\partial_r\eta(r))e^{\eta(r)/2}\over 2} &
- {\delta_\mu^1(\partial_r\eta(r))e^{\eta(r)/2}\over2}
& 0 & -{\delta_\mu^\theta\over r} \\
0 & 0 & {\delta_\mu^\theta\over r} & 0 \\
\end{array}\right),
\end{equation}
where entries in the above matrix are
w.r.t. indices $(\hat{0},\hat{1},\hat{r},\hat{\theta})$.
The  killing spinor equations obtained from non-trivial 
variations of the gravitino are then given by 
\begin{eqnarray} \label{eq:vargrav}
& & (i)
\delta\psi_0=\partial_0\epsilon
-{1\over4}e^{3\eta(r)/2}(\partial_r\eta(r))e^{-\eta(r)}
\gamma^{\hat{r}}\gamma^{\hat{0}}(-1-\gamma^{\hat{0}}\gamma^{\hat{1}})\epsilon
=0;\nonumber\\
& & (ii) \delta\psi_1\sim\delta\psi_0;\nonumber\\
& & (iii)
\delta\psi_r=
\partial_r\epsilon
-{1\over4}\partial_r\eta(r)\gamma^{\hat{0}}\gamma^{\hat{1}}\epsilon=0;
\nonumber\\
& &
(iv)
\delta\psi_\theta
=\partial_\theta\epsilon-{1\over2r}\gamma_{\hat{r}}
\gamma_{\hat{\theta}}\epsilon=0.
\end{eqnarray}
Thus, using
${\gamma_{\hat{r}}\gamma_{\hat{\theta}}\over r}=\gamma^{\hat{2}}
\gamma^{\hat{3}}$
(which has eigenvalues $\pm1$), and
$e^{-\eta(r)}=-{1\over{2\pi}}ln{r\over r_0},$
$a(\theta)={\theta\over2\pi}$,
one sees that (\ref{eq:vargravtheta}) can be consistently solved
for $\epsilon$ to give:
\begin{equation}
\label{eq:eprtheta}
\epsilon(r,\theta)=
e^{-{\eta(r)\over4}}.e^{{1\over2}\gamma^{\hat{2}}\gamma^{\hat{3}}\theta}
\epsilon_0,
\end{equation}
and satisfies 
\begin{equation}
\label{eq:chcondepsS}
(1+\gamma^{\hat{0}}\gamma^{\hat{1}})\epsilon_0=0. 
\end{equation}
Solutions like the one in (\ref{eq:eprtheta}) have been discussed 
eariler in \cite{killsptheta}).
The supersymmetric variation of the dilatino is given by:
\begin{eqnarray}
\label{eq:dilsusytrSst}
& &
\delta\lambda={e^{\eta(r)}\over{4\sqrt{2}}}\biggl[\gamma^{\hat{r}}
\partial_r(e^{-\eta(r)}
\epsilon+i\gamma^{\hat{5}}
\gamma^{\hat{\theta}}\partial_\theta a(\theta))\biggr]\epsilon\nonumber\\
& &
={e^{\eta(r)}\over{4\sqrt{2}}}(1+\gamma^{\hat{0}}\gamma^{\hat{1}})\epsilon=0,
\end{eqnarray}
which, as consistency would require, gives the same chirality
condition
as (\ref{eq:chcondepsS}).
The supersymmetric variation of the gaugino is trivially zero.

\vskip 0.2 true in

(II) \underline{(1,0) U string}

\vskip 0.2 true in

The (1.0) U-string solution is given in equation (\ref{eq:Udef}). 
The supersymmetric variation 
of the dilatino for (\ref{eq:Udef}) is trivially
zero. The variations of the gravitino and gaugino are given by:
\begin{eqnarray}
\label{eq:susytransf}
& & \delta\psi_\mu=\biggl[\bigtriangledown_\mu+{1\over4}
Q_\mu^{\hat{m}\hat{n}}\Gamma^{\hat{m}\hat{n}}\biggr]\epsilon=0,\nonumber\\
& & \delta\chi^{\hat{m}}={1\over\sqrt{2}}
\gamma^\mu P_\mu^{\hat{m}\hat{n}}\Gamma^{\hat{m}}\epsilon=0.
\end{eqnarray}
where $Q_\mu$ and $P_\mu$ were defined earlier. 
The spin connection
$\omega_\mu^{\hat{\mu}\hat{\nu}}$ for (\ref{eq:Udef}) is
given by:
\begin{equation}
\label{eq:spincon}
\omega_\mu^{\hat{\mu}\hat{\nu}}=\left(\begin{array}{cccc}
0 & 0 & 0 & 0 \\
0 & 0 & 0 & 0 \\
0 & 0 & 0 & -{\delta^\theta_\mu\over r}(1-{r\over2}\partial_r\rho(r)) \\
0 & 0 & {\delta^\theta_\mu\over r}(1-{r\over2}\partial_r\rho(r)) & 0
\end{array}\right).
\end{equation}
Now, to evaluate $Q_\mu$ and $P_\mu$ in (\ref{eq:susytransf}),
it is sufficient for our purpose to consider only the $T^2$ that is common to 
$T^4\times T^2$ and $K3\times T^2$. Hence, for (\ref{eq:Udef}), the
matrix $M$ is given by:
\begin{equation}
\label{eq:MU}
M=\left(\begin{array} {cc}
G^{-1} & 0 \\ 
0 & G 
\end{array}\right),
\end{equation}
where $G=EE^T$ is the metric on $T^2$ as given in  (\ref{eq:T2metric}),
for unit determinant (i.e. $\sigma=0$ in (\ref{eq:T2metric})) 
with
\begin{equation}
\label{eq:Edef}
E=\left(\begin{array}{cc}
-e^{-{\rho\over2}} & ce^{\rho\over2} \\
0 & -e^{\rho\over2}
\end{array}\right).
\end{equation}
Thus, one gets:
\begin{eqnarray}
\label{eq:Qdef}
& & Q_\mu^{\hat{m}\hat{n}}
=(V_RL\partial_\mu
V_R^T)^{\hat{m}\hat{n}}
={1\over2}(E^{-1}\partial_\mu E+E^T\partial_\mu
E^{T\ -1})^{\hat{m}\hat{n}}
\nonumber\\
& & 
=\left(\begin{array}{cc}
0 & -{e^\rho\over2}(\partial_\mu c)\delta^\theta_\mu
\\
{e^\rho\over2}(\partial_\mu c)\delta^\theta_\mu & 0 
\end{array}\right),
\end{eqnarray}
with matrix components denote indices $\hat{m},\hat{n}$.
Similarly,
\begin{eqnarray}
\label{eq:Pdef}
& & \gamma^\mu P_\mu
^{\hat{m}\hat{n}}
=(\gamma^\mu V_LL\partial_\mu V_R^T)^{\hat{m}\hat{n}}
={1\over2}\gamma^\mu(E^{-1}\partial_\mu E-E^T\partial_\mu
E^{T\ -1})^{\hat{m}\hat{n}}
\nonumber\\
& & ={1\over2}\left(\begin{array} {cc}
-\gamma^{\hat{r}}\partial_r\rho(r) &
-e^{\rho(r)}\gamma^{\hat{\theta}}\partial_\theta c(\theta) \\
-e^{\rho(r)}\gamma^{\hat{\theta}}\partial_\theta c(\theta) &
\gamma^{\hat{r}}\partial_r\rho(r)
\end{array}\right).
\end{eqnarray}
The only non-trivial variation of the gravitino is given for
$\mu=\theta$:
\begin{eqnarray} 
\label{eq:vargravtheta}
& &
\delta\psi_\theta=\biggl(\partial_\theta+{1\over2}\omega_\theta^{\hat{r}
\hat{\theta}}\Gamma_{\hat{r}}\Gamma_{\hat{\theta}}
+{1\over2}Q_\theta^{\hat{4}\hat{5}}\Gamma^{\hat{4}}\Gamma^{\hat{5}}\biggr)
\epsilon\nonumber\\
& & =
(\partial_\theta-{\gamma_{\hat{r}}\gamma_{\hat{\theta}}\over2r})\epsilon
+{1\over4}\partial_r\rho(r)\gamma_{\hat{r}}\gamma_{\hat{\theta}}\epsilon
-{1\over4}(\partial_\theta c(\theta))e^{\rho(r)}
\Gamma^{\hat{4}}\Gamma^{\hat{5}}\epsilon=0.
\end{eqnarray}
Using
${\gamma_{\hat{r}}\gamma_{\hat{\theta}}\over r}=
\gamma^{\hat{2}}\gamma^{\hat{3}}$ 
(which has eigenvalues $\pm1$) and
$e^{-\rho(r)}=-{1\over{2\pi}}ln{r\over r_0},$
$c(\theta)={\theta\over2\pi}$,
one sees that (\ref{eq:vargravtheta}) can be consistently solved
for $\epsilon$ by imposing:
\begin{equation}
\label{eq:solepstheta1}
\partial_\theta\epsilon-{1\over2}\gamma^{\hat{2}}\gamma^{\hat{3}}\epsilon=0,
\end{equation}
and
\begin{equation}
\label{eq:susycoomd1}
{1\over4}\partial_r\rho(r)\gamma_{\hat{r}}
\gamma_{\hat{\theta}}\epsilon
-{1\over4}(\partial_\theta c(\theta))e^{\rho(r)}
\Gamma^{\hat{4}}\Gamma^{\hat{5}}\epsilon=0.
\end{equation}
Equation (\ref{eq:solepstheta1}) has a solution:
\begin{equation}
\label{eq:solepstheta2}
\epsilon(\theta)=e^{{\gamma^{\hat{2}}\gamma^{\hat{3}}\over2}\theta}\epsilon_0,
\end{equation}
and (\ref{eq:susycoomd1}) gives:
\begin{equation}
\label{eq:susycoomd2}
(1+\gamma^{\hat{2}}\gamma^{\hat{3}}\Gamma^{\hat{4}}\Gamma^{\hat{5}})\epsilon_0=0.
\end{equation}
One can show that one gets identical  condition as
(\ref{eq:susycoomd2}) from the variation of the
gaugino. Equation
(\ref{eq:susycoomd2}) again
implies that 1/2 of spacetime supersymmetry is preserved.

\vskip 0.2 true in

(III) \underline{(1,0) T string}

\vskip 0.2 true in

Let us consider the T string solution of (\ref{eq:Tdef}). For this case,
the matrix $M$ has the general O(2,2) form:
\begin{equation}
\label{eq:Mdef}
M=\left(\begin{array} {cc}
G^{-1} &  -G^{-1}B \\
BG^{-1}&  G-BG^{-1}B 
\end{array}\right),\end{equation}
which
implies:
\begin{equation}
\label{eq:VRLdef}
V=\left(\begin{array} {c} 
V_R \\
V_L \end{array}\right)
={1\over\sqrt{2}}\left(\begin{array}{cc}
E^{-1} &  E^T-E^{-1}B \\
E^{-1} & -E^T-E^{-1}B
\end{array}\right).
\end{equation}
The $T^2$ metric is given by:
$G=e^{-\sigma(r)}{\bf 1}_2=EE^T$,
which implies
$E=e^{-{\sigma(r)\over2}}{\bf 1}_2$.
$Q_\mu$ and $P_\mu$ are then given by:
\begin{eqnarray}
\label{eq:Qdef2}
& & Q_\mu
^{\hat{m}\hat{n}}
={1\over2}\biggl(E^{-1}\partial_\mu E
+ E^T\partial_\mu E^{T\ -1} + E^{-1}(\partial_\mu B)
E^{T\ -1}\biggr)^{\hat{m}\hat{n}}
\nonumber\\
& & = {1\over2}e^{\sigma(r)}\left(\begin{array}{cc}
0 & \partial_\mu b\\
-\partial_\mu b & 0 
\end{array}\right)\delta^\theta_\mu,
\end{eqnarray}
and
\begin{eqnarray}
\label{eq:Pdef2}
& & \gamma^\mu P_\mu^{\hat{m}\hat{n}}
=  {1\over2}\gamma^\mu\biggl(E^{-1}\partial_\mu
E+E^T\partial_\mu E^{T\ -1}-E^{-1}(\partial_\mu
B)E^{T\ -1}\biggr)^{\hat{m}\hat{n}}
\nonumber\\
& & ={1\over2}
\left(\begin{array}{cc}
-\gamma^{\hat{r}}
\partial_r\sigma(r) & e^{\sigma(r)}\gamma^{\hat{\theta}}\partial_\theta b\\
-e^{\sigma(r)}\gamma^{\hat{\theta}}\partial_\theta b & 
-\gamma^{\hat{r}}
\partial_r\sigma(r)
\end{array}\right)\delta^\theta_\mu.
\end{eqnarray}
The analysis parallels that of the U string with
$\rho\rightarrow\sigma$ and $c\rightarrow-b$. Thus, the 
supersymmetry condition is
\begin{equation}
\label{eq:susycoomd22}
(1-\gamma^{\hat{2}}\gamma^{\hat{3}}\Gamma^{\hat{4}}\Gamma^{\hat{5}})
\epsilon_0=0.
\end{equation}
Again, this condition is the same as the one obtained by the 
supersymmetric variation of the gaugino.

We have therefore obtained the supersymmetry properties of 
the string solution\cite{duffetal} of the heterotic string
theory, referred to here as $(1, 0)$-solutions. The $(p, q)$-
solutions can then be generated by applying the $SL(2, Z)_H$
discussed earlier. In the present case, as mentioned earlier,
the role of this $SL(2, Z)$ is played by constant coordinate
transformations acting on coordinates $x^5$ and $x^6$. These transformations
generate other classical solutions, which in fact are
equivalent to the original ones as they are generated by 
constant coordinate transformations. Also, as in the 
case of ten-dimensional type IIB theory, the string tension
is once again given by an $SL(2, Z)$ invariant function,
from which one can read tensions of strings with 
different quantum numbers. This expression involves moduli
as well as $(p,q)$ quantum numbers denoting charges.
As we will see shortly, the supersymmetry of the transformed 
solutions are in general  different than the original ones. 
As a result, when one combines many such string to 
construct a network configuration, one has a situation  
where individual strings contribute differently to the
total energy and supersymmetry is broken to 1/4. 

Now, it is clear that $S$-string mentioned earlier is
invariant under the above $SL(2, Z)$. As a result, 
one does not have any new supersymmetry condition 
arising out of their action on such solutions.
In fact, we already notice a crucial difference among supersymmetry
conditions of $S-, T-$ and $U$ - strings given in equations
(\ref{eq:chcondepsS}), (\ref{eq:susycoomd2}) and
(\ref{eq:susycoomd22}),
namely that, in addition to the space-time gamma matrices that appear
in equations for $S$-string, we also have internal gamma matrices
appearing in the $T$- and $U$- string supersymmetry condition.
They finally allow one to make an alignment using space-time and
internal orientations of strings in order to construct string
networks.

Now, to obtain the killing spinor for
 ``(p,q)'' $U$ or $T$ string, 
one performs a constant coordinate SL(2)$_H$ transformation on 
the killing spinors in equations (\ref{eq:susycoomd2}) 
and (\ref{eq:susycoomd22}). The effect of 
SL(2,Z) S-duality transformations in heterotic string and 
D=10 IIB theories have been studied \cite{ortin,sen97}.
It is known that spinors transform as a representation of the maximal compact
subgroup which in this case truns out to be parameterized by 
a rotation angle. In this case, this 
corresponds to a rotation induced by constant
coordinate transformation in $(x^5,x^6)$-space.
Then the new
supersymmetry condition for the ``(p,q)'' $U$ or
$T$ string will be:
\begin{equation}
\label{eq:susycoomdSL2}
\biggl[1\pm\gamma^{\hat{2}}\gamma^{\hat{3}}\Gamma^{\hat{4}}
({\rm cos}\ \theta\Gamma^{\hat{5}}+{\rm sin}\ 
\theta\Gamma^{\hat{6}})\biggr]\epsilon=0,
\end{equation}
($\pm$ for $U/T$ ``(p,q)'' strings) where using \cite{Schwarz,ortin},
\begin{equation}
\label{eq:thetadef}
e^{i\theta}={p-q\tau^*\over{|p-q\tau^*|}},
\end{equation}
where $\tau$ is the complex structure generated from 
metric components $G_{55, 56, 66}$.

We now discuss the construction of 
string networks for the above string solutions.
First, as already stated, S-string remains unchanged under SL(2,Z)$_H$.
Moreover, since supersymmetry condition (\ref{eq:chcondepsS}) is
also invariant under SL(2,Z)$_H$, one does not have any possibility
to put several different types of strings together to form a network.
These can however be formed using multiplets of $T$ and $U$ strings.
To see this from the point of view of supersymmetry, we rewrite
(\ref{eq:susycoomdSL2}) by defing spinors which are eigenvectors
of $i\Gamma^{\hat{4}}\Gamma^{\hat{5}}$,
denoted by $\epsilon_\pm$. Choosing the following
representation for the six-dimensional $\Gamma^{\hat{m}}$'s:
\begin{equation}
\label{eq:intGammadef}
\Gamma^{\hat{4},\hat{5},\hat{6}}=\sigma^{2,1,3}\otimes
{\bf 1}_2\otimes{\bf 1}_2,
\end{equation}
$\epsilon$ can be taken to be:
\begin{equation}
\label{eq:epspmdef1}
\epsilon=\left(\begin{array}{c}
\tilde\epsilon_+ \\
\tilde\epsilon_-
\end{array}\right)
=\left(\begin{array}{c}
\epsilon_+\otimes\epsilon_0\\
\epsilon_-\otimes\epsilon_0
\end{array}\right),
\end{equation}
where $\tilde\epsilon_\pm$, $\epsilon_\pm$ and $\epsilon_0$ 
are 16-, 8-, and 2-component spinors respectively, where.
\begin{equation}
\label{eq:epspmd2}
i\Gamma^{\hat{4}}\Gamma^{\hat{5}}\epsilon_\pm=\pm\epsilon_\pm.
\end{equation}
Hence, one gets the following result from (\ref{eq:susycoomdSL2}):
\begin{equation}
\label{eq:SL2epspq}
\epsilon_++i\epsilon_-=\pm
i e^{-i\theta}\gamma^{\hat{2}}\gamma^{\hat{3}}
(\epsilon_+-i\epsilon_-).
\end{equation}
Similar condition can be written for $T$-string as well.
The 1/4 spacetime supersymmetry of the $U$ and $T$ string networks can
now be established by following the arguments of \cite{sen97,bhatt}.

To summarize, we have now shown that by superimposing $T$ and 
$U$ string networks of the heterotic theory, one can have network 
configuration in this theory preserving only $1/4$ of the 
supersymmetry. We have therefore verified a prediction of 
string-triality from the point of view of string network 
construction in heterotic theory. 

We now discuss the existence of $S$ and $T$ strings in type IIB
on $K3\times T^2$, as prediced by the string-triality. 
First, to see the existence of $S$-string networks, we notice that
the four dimensional $S$-strings are charged under the four 
dimensional axion-dilaton fields. Among these, the axion 
field orginates from the NS-NS 2-form $B_{M N}$ in ten dimensions
and the dilaton comes from the ten-dimensional dilaton. As a result,
the $S$ string solution mentioned earlier for IIB case, can 
in fact be identified with the $(1, 0)$-string of the ten-dimensional 
theory. Moreover due to these identifications, the $S$-duality 
in four dimensions is seen to follow from the one in $D=10$. As
a result, the existence of the $S$-string networks in IIB theory  also 
follows from the $(p, q)$-string networks in ten-dimensions. 
Similar arguments hold for the $S$-string networks in type IIA as well.
However one now has to use the fact that they are equivalent to
IIB theories, when compactified on a circle.

To complete the arguments in favor of the string-triality, 
we now discuss the existence of $T$-string networks in IIB theories 
and $U$-string networks in IIA theories. 
By using the mapping between the two theories, when 
compactified on a circle, we automatically have the existence 
of the string network in IIA, given the one for IIB. 
So, finally to show the presence of $T$-string networks in 
IIB theories, we notice that, by promoting the solution
to six dimensions, and then by using Hodge duality
of 3-form field strengths, one obtains a `smeared' 
or delocalized string solution in six dimensions. Promoting this further to 
a smeared solution in ten dimensions, one again finds the 
string networks in these theories. In other words, to put
$S$-string solutions to an appropriate
form in $D=10$ we needed to use
the Hodge-duality in four dimensions, whereas in the present
case we make use of it in six dimensions. In both these
cases we ignore the $K3$ factor, by writing down the 
smeared solutions with no coordinate dependence along 
$K3$ directions. 

To conclude, we have given evidence in favor of string-triality 
in four dimensions by explicitly examining the supersymmety
properties in the heterotic theory. We have also given  arguments
in favor of the existence of string-networks in type II
theories, that are predicted by the above triality. The results
are summarized in Fig 1.

\begin{figure}[p]
\centerline{\mbox{\psfig{file=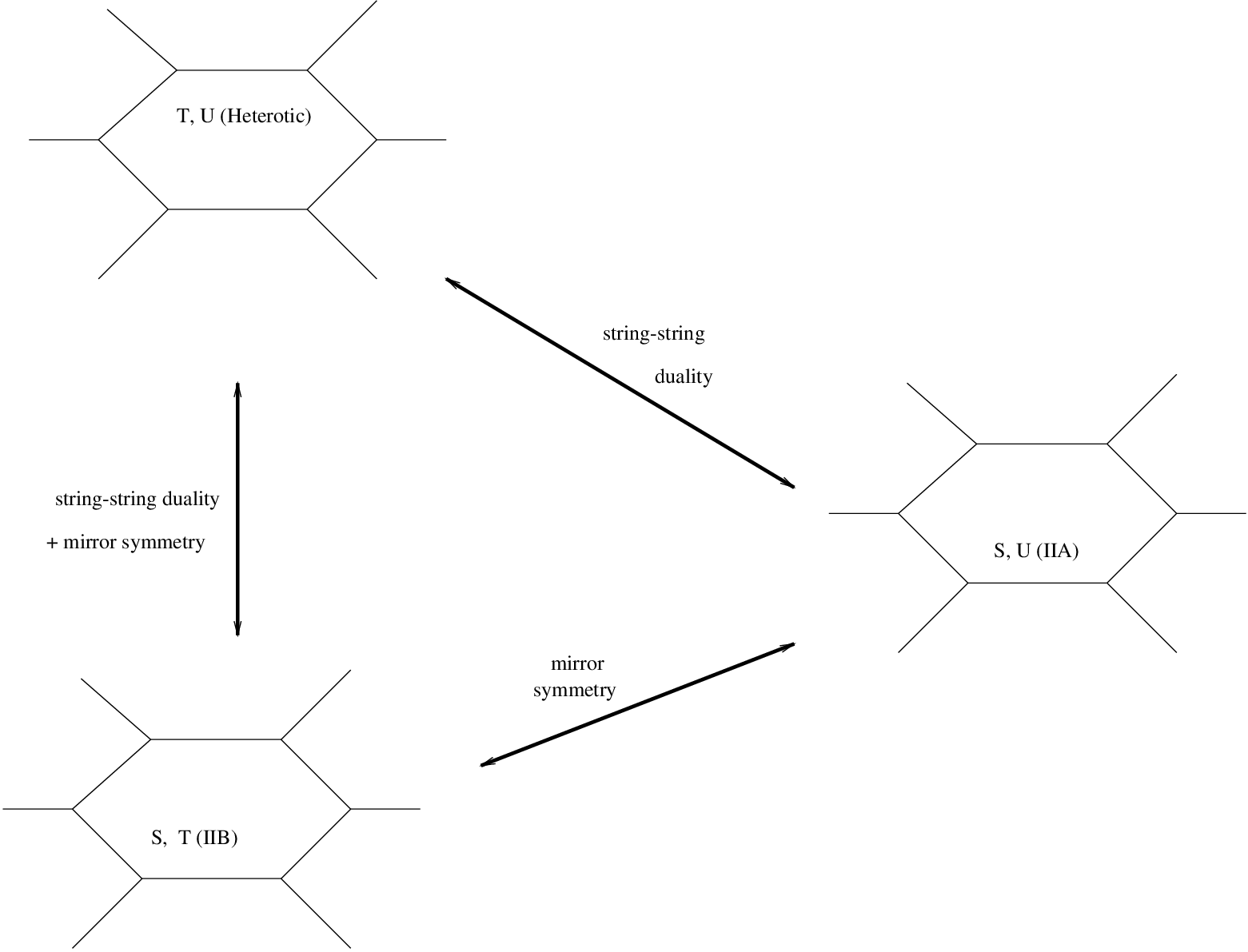,width=0.8\textwidth}}}
\caption{Triality of string networks in four dimensions}
\end{figure}


\begin{thebibliography}{99}
\newcommand{\np}{{\it Nucl. Phys.} {\bf B}}
\newcommand{\pl}{{\it Phys. Lett.} {\bf B}}
\newcommand{\prd}{{\it Phys. Rev. }{\bf D}}
\newcommand{\prl}{{\it Phys. Rev. Lett.}}
\newcommand{\mpl}{{\it Mod. Phys. Lett. }{\bf A}}
\newcommand{\ijmp}{{\it Int. J. Mod. Phys.}{\bf A}}
\newcommand{\cqg}{\it Class. Quant. Grav.}
\newcommand{\jmp}{\it J. Math Phys.}
\newcommand{\cmp}{\it Comm. Math. Phys.}
\newcommand{\pr}{\it Phys. Rept}
\bibitem{duffetal}  M.J. Duff, James T. Liu, J. Rahmfeld, 
{\np}{\bf 459} (1996) 125, [hep-th/9508094].
\bibitem{Luloop} 
M.~J. Duff and J.~X. Lu,
\newblock {\sl Loop
expansions and string/five-brane duality},   
{\np} {\bf 357} (1991) 534, [hep-th/9305142].

 
\bibitem{Khurifour}M.~J. Duff and R.~R. Khuri,
\newblock {\sl Four-dimensional string/string duality},
{\np}{\bf 411} (1994) 473,  [hep-th/9305142].

\bibitem{Lublack}
M.~J. Duff and J.~X. Lu,
\newblock {\sl Black and super $p$-branes in diverse dimensions},
{\np}{\bf 416} (1994) 301, [hep-th/9306052].
 
\bibitem{Minasian} 
M.~J. Duff and R.~Minasian,
\newblock {\sl Putting string/string duality to the test},
{\np}{\bf 436} (1995) 507,  [hep-th/9406198].

\bibitem{Duffclassical}
M.~J. Duff,
\newblock {\sl Classical/quantum duality},
\newblock in {\em Proceedings of the International
High Energy Physics Conference, Glasgow} (July 1994),
(Eds. Bussey and Knowles).

\bibitem{Khuristring}
M.~J. Duff, R.~R. Khuri and J.~X. Lu,  
\newblock {\sl String solitons},
{\pr}{\bf 259} (1995) 213, [hep-th/9412184].

\bibitem{Duffstrong}M.~J. Duff,
\newblock {\sl Strong/weak coupling duality from the dual string},
{\np}{\bf 442} (1995) 47 [hep-th/9501030].

\bibitem{Hull}
C.~M. Hull and P.~K. Townsend,
\newblock {\sl Unity of superstring dualities}, {\np} {\bf 438}
(1995) 109, [hep-th/9410167].


\bibitem{Witten} E.~Witten, 
\newblock {\sl String theory dynamics in various dimensions},
{\np}{\bf 443} (1995) 85, [hep-th/9503124].


\bibitem{Senssd}
A.~Sen,
\newblock {\sl String string duality conjecture in six dimensions and
charged solitonic strings}, {\np}{\bf 450}
(1995) 103, [hep-th/9504027].

\bibitem{Harvey}
J.~A. Harvey and A.~Strominger,
\newblock {\sl The heterotic string is a soliton},
{\np}{\bf 449} (1995) 535,  [hep-th/9504047].



\bibitem{Dine}M.~Dine, P.~Huet and N.~Seiberg,
\newblock {\sl Large and small radius in string theory},
{\np} {\bf 322} (1989) 301.

\bibitem{kehagias}
  A. Kehagias, 
\newblock {\sl N=2 Heterotic Stringy Cosmic Strings}, TUM-HEP-262/96,
[hep-th/9611110].

\bibitem{ortin} T. Ortin,
\newblock {\sl Sl(2,R)-duality covariance
of killing spinors in axion-dilaton black
holes}, {\prd}{\bf 51}
(1995) 790, [hep-th/9404035].

\bibitem{youm} D. Youm, 
\newblock {\sl Black Holes and Solitons in String Theory},
{\pr} {\bf 316} (1999) 1, [hep-th/9710046].

\bibitem{schwarz} J. Schwarz, {\it Lectures
on Superstring and M
Theory Dualities}, [hep-th/9607201].

\bibitem{aha} O. Aharony, J. Sonnenschein and S. Yankielowicz, 
\newblock {\sl Interactions of strings and D-branes from M
theoryInteractions of strings and D-branes from M theory},
{\np}{\bf 474} (1996) 309 [hep-th/9603009]; O. Aharony and A. Hanany,
\newblock {\sl Branes, Superpotentials and Superconformal Fixed
Points}, {\np}{\bf 504} (1997) 239, [hep-th/9704170];
O. Aharony, A. Hanany and B. Kol, 
\newblock 
{\sl Webs of (p,q) 5-branes, Five Dimensional Field Theories and
Grid Diagrams}
JHEP {\bf 9801}(1998) 002, [hep-th/9710116].

\bibitem{dasg} K. Dasgupta and S. Mukhi, 
\newblock {\sl BPS Nature of 3-String Junctions}
{\pl}{\bf 423} (1998) 261, [hep-th/9711094].


\bibitem{rey} S-J. Rey and J-T. Yee, 
\newblock {\sl BPS Dynamics of Triple (p,q) String
Junction}, {\np}{\bf 526} (1998) 229,
[hep-th/9711202].

\bibitem{sen97} A. Sen, {\it String Network}, {\bf JHEP 9803:005} 
(1998), [hep-th/9711130].

\bibitem{bhatt} S. Bhattacharyya, A. Kumar and S. Mukhopadhyay,
\newblock {\sl String Network and U-Duality},
{\prl} {\bf 81} (1998) 754, [hep-th/9801141].

\bibitem{lee} M. Krogh and S. Lee, 
\newblock {\sl String Network from M-theory},
{\np}{\bf 516} (1998) 241,  
[hep-th/9712050]; Y. Matsuo and K. Okuyama, 
\newblock
{\sl BPS Condition of String Junction from M theory},
{\pl}{\bf 426} (1998) 294, [hep-th/9712070].

\bibitem{thorla} C. Callan and
L. Thorlacious,
\newblock {\sl Worldsheet Dynamics of String
Junctions},
{\np}{\bf 534} (1998) 121, [hep-th/9803097].

\bibitem{sudipt} S. Mukherji,
\newblock {\sl On the SL(2,Z) Covariant World-Sheet Action with Sources},
{\mpl} {\bf 13} (1998) 2819, [hep-th/9805031].



\bibitem{subir}A. Kumar and S. Mukhopadhyay,  
\newblock {\sl U-duality and Network Configurations of Branes},
{\ijmp}{\bf 14} (1999)
3252, [hep-th/9806126].

\bibitem{zwieb} M. Gabardiel and B. Zwiebach,
\newblock {\sl Exceptional groups from open
strings},
{\np}{\bf 518} (1998) 151, [hep-th/9709013];
M. Gabardiel, T. Hauer and B. Zwiebach,
\newblock {\sl Open string - string junction transitionsOpen string -
string junction transitions},
{\np}{\bf 525} (1998) 117, [hep-th/9801205];
O. Bergman, 
\newblock{\sl Three-Pronged Strings and 1/4 BPS States in N=4
Super-Yang-Mills Theory},
{\np}{\bf 525} (1998) 104,
[hep-th/9712211]; O. Bergman and A. Fayyazuddin, 
\newblock {\sl String Junctions and BPS States in Seiberg-Witten Theory}
{\np}{\bf 531} (1998) 108, [hep-th/9802033];
O. Bergman and B. Kol, 
\newblock {\sl String Webs and 1/4 BPS Monopoles}
{\np}{\bf 536} (1998) 149, [hep-th/9804160];
K. Hashimoto, H. Hata and N. Sasakura,   
\newblock{\sl 3-String Junction and BPS Saturated Solutions in SU(3) Supersymmetric Yang-Mills
Theory}
{\pl}{\bf 431} (1998) 303, [hep-th/9803127]; 
\newblock{\sl 
Multi-Pronged Strings and BPS Saturated Solutions in SU(N) Supersymmetric
Yang-Mills Theory}
{\np}{\bf 535} (1998) 83,
[hetp-th/9804164];
P. Ramadevi, 
\newblock{\sl Supergravity Solution for Three-String Junction in
  M-Theory},
JHEP 0006 (2000) 005, [hep-th/9906247]; A. Kumar,
\newblock {\sl Charged Macroscopic type II Strings and their Networks},
JHEP 9912 (1999) 001; A. Kumar,
\newblock {\sl Non-Planar String Networks on Tori},
JHEP 0003 (2000) 010, [hep-th/0002150]; 
A. Kumar and S. Mukherji, 
\newblock {\sl On Charged Strings and their Networks }
[hep-th/0005093];
(For a more comprehensive list of references, see) B. Kol, 
\newblock {\sl On the Spatial Structure of Monopoles},
[hep-th/0002118].

\bibitem{Schwarz}  J. Schwarz, 
\newblock{\sl An SL(2,Z) Multiplet of Type IIB Superstrings},
Phys.Lett. B{\bf 360} (1995) 13-18; Erratum-ibid. B{\bf 364} (1995) 252
[hep-th/9508143].
\newblock{\sl}
\bibitem{killsptheta}
H. Lu, C.N. Pope, J. Rahmfeld, 
\newblock{\sl A Construction of Killing Spinors on $S^n$},
J.Math.Phys. 40 (1999) 4518-4526, 
[hep-th/9805151]; P. M. Llatas, J. M. Sanchez de Santos, 
\newblock{\sl S-Duality,
SL(2,Z) Multiplets and Killing Spinors}, Phys.Lett. B484 (2000) 306-314,
[hep-th/9912159].

\end{thebibliography}
\end{document}